\documentclass[floats,aps,prb,twocolumn]{revtex4}
\usepackage{graphicx}
\usepackage{amsmath}
\newcommand{\rr}{{\bf r}}
\newcommand{\qq}{{\bf q}}

\newcommand{\RR}{{\bf R}}
\newcommand{\kk}{{\bf k}}
\newcommand{\pp}{{\bf p}}

\begin{document}

\title{Many-Impurity Effects in Fourier Transform Scanning Tunneling
Spectroscopy} \author{O. Kodra and W. A. Atkinson} 
\affiliation{Department of Physics, Trent University, 1600 West Bank Dr.,
Peterborough ON, K9J 7B8, Canada}
\date{\today}

\begin{abstract}
Fourier transform scanning tunneling spectroscopy (FTSTS) is a useful
technique for extracting details of the momentum-resolved electronic
band structure from inhomogeneities in the local density of states due
to disorder-related quasiparticle scattering.  To a large extent,
current understanding of FTSTS is based on models of Friedel
oscillations near isolated impurities. Here, a
framework for understanding many-impurity effects is developed based
on a systematic treatment of the variance $\Delta \rho^2(\qq,\omega)$
of the Fourier transformed local density of states $\rho(\qq,\omega)$.
One important consequence of this work is a demonstration that the
poor signal-to-noise ratio inherent in $\rho(\qq,\omega)$ due to
randomness in impurity positions can be eliminated by configuration
averaging $\Delta \rho^2(\qq,\omega)$.  Furthermore, we develop a
diagrammatic perturbation theory for $\Delta \rho^2(\qq,\omega)$ and
show that an important bulk quantity, the mean-free-path, can be
extracted from FTSTS experiments.
\end{abstract}

\pacs{71.18.+y,71.55.-i,71.55.Jv,73.20.-r,74.50.+r,74.81.-g}

\maketitle

\section{Introduction}
As a result of technical advances in scanning tunneling microscopy
(STM) methods, it has become possible to measure the quasiparticle
spectrum at surfaces of a variety of materials over large areas with
both atomic spatial resolution and mV energy resolution.  While STM is
a powerful and direct probe of electronic wavefunctions, the
interpretation of STM spectra generally requires comparison
with a model spectrum calculated using approximate many-body
techniques (for example, the local density approximation).  This
can be problematic in situations where details of the
surface are poorly known, or where traditional theoretical methods
fail, as in the high temperature superconductors.  The question of
how one can usefully analyze STM spectral maps without a detailed
quantitative model of the electronic structure is therefore paramount.

\begin{figure}[tb]
\includegraphics[width=\columnwidth]{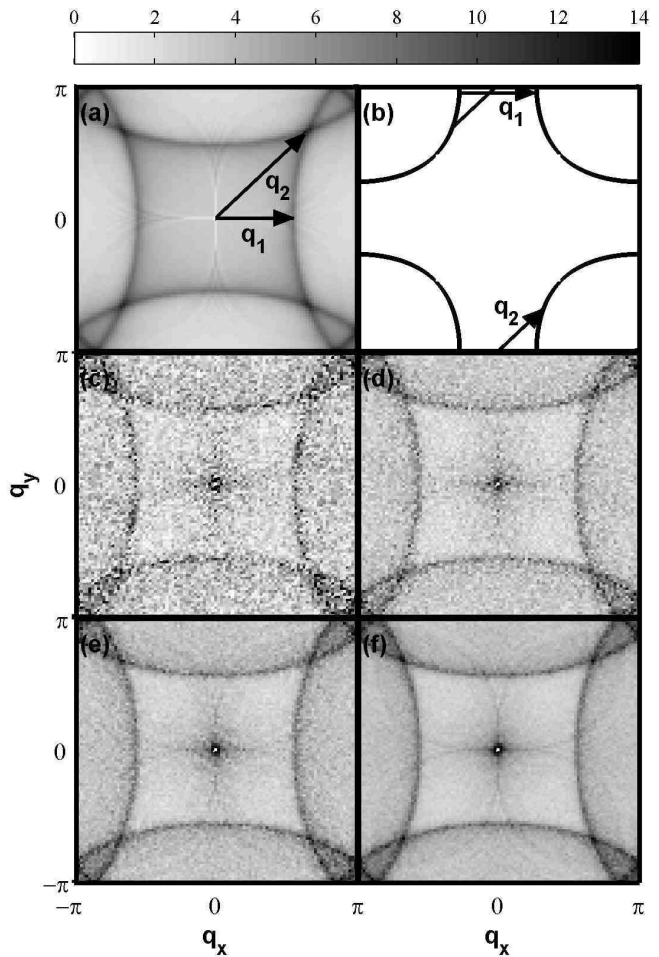}
\caption{(color online) Effect of randomness in the impurity
distribution on the FTDOS of the 2D tight-binding metal described in
Sec.~\protect\ref{numerics}.  (a) $\rho(\qq,\omega)$ is plotted for a
single pointlike impurity embedded in a $100\times 100$ lattice.  The
Fermi surface contour $\epsilon_\kk=0$ is shown in (b).  Arrows
$\qq_1$ and $\qq_2$ show the scattering transitions producing the
features labelled $\qq_1$ and $\qq_2$ in (a). The FTDOS for 3\%
randomly distributed impurities is shown averaged over
$N_\textrm{cfg}$ configurations for (c) $N_\textrm{cfg}=1$, (d)
$N_\textrm{cfg}=4$, (e) $N_\textrm{cfg}=10$, (f) $N_\textrm{cfg}=29$.
In (c), $|\rho(\qq,\omega)|$ is plotted, while in (d)-(f), we plot the
root-mean-square FTDOS $\rho_\textrm{rms}(\qq,\omega) \equiv
\sqrt{\langle \Delta \rho^2(\qq,\omega) \rangle_{N_\textrm{cfg}}}$.
In all figures, we take $\omega = 0$ and an impurity potential of
$V=10t_1$.  Other model details are given in
Sec.~\protect\ref{numerics}.  }
\label{fig1}
\end{figure}

Recently, it was shown that details of the quasiparticle dispersion
can be extracted from the spatial Fourier transform of 
STM spectra on disordered surfaces.\cite{sprunger}  
The technique is often referred to as Fourier transform
scanning tunneling spectroscopy (FTSTS) and requires that maps of the
local density of states (LDOS) 
$\rho(\rr,\omega)$ at energy $\omega$ be made over reasonably 
large surface areas, from which the Fourier  transform,
$\rho(\qq,\omega) = \sum_\rr \rho(\rr,\omega) \exp(-i\qq\cdot\rr)$, 
can be obtained. In the simplest case, that of noninteracting
quasiparticles, the LDOS at position $\rr$ and energy $\omega$ is
\begin{equation}
\rho(\rr,\omega) =  \sum_n
|\Psi_n(\rr)|^2\delta(\omega-E_n),
\label{exactldos}
\end{equation}
where $\Psi_n(\rr)$ is the $n^\textrm{th}$ quasiparticle eigenstate
with energy $E_n$.  For a clean surface, $\Psi_n(\rr)$ are Bloch
states of the form $\phi_{j,\kk}(\rr)e^{i\kk\cdot\rr}$ with wavevector
$\kk$ and band index $j$, and $\rho(\rr,\omega)=\sum_{j,\kk}
|\phi_{j,\kk}(\rr)|^2\delta(\omega-E_{j\kk})$ has only the spatial
periodic modulations of the unit cell.  However, in the presence of
defects, the eigenstates contain mixtures of Bloch states with
different values of $\kk$, and the probability density
$|\Psi_n(\rr)|^2$ exhibits standing wave patterns (Friedel
oscillations) arising from the interference between these different
$\kk$-values.  Near an isolated impurity, for example, one can
decompose the eigenstates at energy $\omega$ into initial and final
(scattered) states with wavevectors $\kk_i$ and $\kk_f$
respectively. Then, the Fourier transformed density of states (FTDOS)
is peaked at wavevectors $\qq=\kk_f-\kk_i$ for which $\kk_f$ and
$\kk_i$ lie on constant energy contours of energy $\omega$.  This is
illustrated in Fig.~\ref{fig1}: in panel (a) the FTDOS is calculated
(details given later) at the Fermi energy $\omega=0$ for a single
impurity in a 2D tight-binding metal whose Fermi surface is shown in
panel (b).  The arrows $\qq_1$ and $\qq_2$ in (b) label transitions
between states on the Fermi surface which contribute to the maxima at
$\qq_1$ and $\qq_2$ in (a).  There is, therefore, a direct connection
between $\rho(\qq,\omega)$ and the quasiparticle energy dispersion
$\epsilon_\kk$.  The possibility of using FTSTS to map out $\epsilon_\kk$
was first demonstrated by Sprunger {\em et
al.}\cite{sprunger} and FTSTS has since been used by several groups to study
metal surfaces,\cite{sprunger2,hansen}
semiconductors,\cite{morgenstern,ruffieux} semimetals,\cite{vonau} and
strongly-correlated
superconductors.\cite{davis,kapitulnik,yazdani,lupien} Quite recently,
an extension of this technique to study inelastic scattering in
strongly-correlated superconductors has been proposed.\cite{iftsts}

The usefulness of the FTSTS method, however, is limited by the large
signal-to-noise ratio which is invariably present.  This is an
intrinsic problem, as we now illustrate, which arises from the
randomness of the impurity positions.  We assume, for simplicity, that
a random distribution of $N_i$ identical impurities sit at sites
$\RR_i$, $i=1,\ldots,N_i$ and that the impurities do not interfere
substantially, so that each impurity modifies the LDOS by an amount
$\delta\rho(\rr-\RR_i,\omega)$.  This term is meant to describe the
Friedel oscillations in the vicinity of an isolated impurity, and its
Fourier transform $\delta\rho(\qq,\omega)$ is the quantity one wishes
to study experimentally.  For simplicity, we will ignore sub-unit-cell
modulations of the LDOS due to the crystalline potential so that the
LDOS is uniform in the absence of impurities.  Then, with the
impurities present, $\rho(\rr,\omega) =
\rho_0(\omega)+\sum_{i=1}^{N_i}\delta\rho(\rr-\RR_i,\omega)$, and
it follows that
\begin{equation}
\rho(\qq,\omega) =
[N\rho_0(\omega)+N_i\delta\rho(\qq,\omega)]\delta_{\qq,0} + \delta
\rho(\qq,\omega) \sqrt{N_i} f(\qq),
\label{ftdos1}
\end{equation}
where $\delta \rho(\qq,\omega) = \sum_{\rr} e^{-i\qq\cdot \rr}
\delta \rho(\rr,\omega)$ is the Fourier transform of the scattering
pattern due to a {\em single} impurity, $N$ is the number of spatial points
included in the Fourier transform, and 
\[
f(\qq) = \frac{ \delta_{\qq\neq 0} }
%\frac{1}
{\sqrt{N_i}}
\sum_{i=1}^{N_i} e^{-i\qq\cdot\RR_i}, 
\]
 is a random function of $\qq$ normalized such that $|f(\qq)|\sim 1$
and $\delta_{\qq\neq 0} = 1-\delta_{\qq,0}$.  The form of
Eq.~(\ref{ftdos1}) was first demonstrated in the limit of weak
impurities by Capriotti {\em et al.}\cite{capriotti} and then for a
dilute distribution of pointlike impurities of arbitrary strength by
Zhu {\em et al.}\cite{ZAH}.  We make two comments regarding
Eq.~(\ref{ftdos1}). First, for a {\em particular} disorder
configuration, and for $\qq\neq 0$, $\rho(\qq,\omega)$ is the product
of a useful signal $\delta \rho(\qq,\omega)$ and a noise term
$f(\qq)$.  In practice, this means that the noise is as large as the
signal, and that it can be nearly impossible to resolve the fine
details of the FTDOS spectrum which have been predicted in numerous
calculations.  This is particularly true in the cuprate
superconductors, which are inherently dirty materials. This point is
illustrated in Fig.~\ref{fig1} where the FTDOS is shown for a 2D metal
for both a single point scatterer in panel (a), and a random
configuration of 3\% impurities in panel (c) (the details of the
calculation will be given below). The figure clearly demonstrates the
difficulty one faces when dealing with disordered systems; while the
FTDOS pattern is visible in (c), one would have difficulty making a
quantitative comparison of the structure with (a).  The second comment
is that in nearly all calculations to date, $\delta\rho(\qq,\omega)$
has been calcuated in the clean limit, ie.\ for a single impurity.
This approximation captures many qualitative features of the spectrum,
but also risks failing when the scattering rate is large, as appears
to be the case in the cuprate superconductors.  It is natural to
assume that disorder, in addition to producing noise, must introduce a
quasiparticle scattering rate $\gamma$ which broadens $\delta
\rho(\qq,\omega)$, but it is in fact not obvious.  Formally, the
elastic scattering rate appears when one averages a large number of
impurity configurations, and is not defined when one studies a
particular configuration of impurities.  If, for example, we try to
take the disorder average $\langle \rho(\qq,\omega)\rangle$, where
$\langle\ldots\rangle$ indicates a configuration average, we find that
since $f(\qq)$ is a sum of random phases, $\langle f(\qq)\rangle = 0$
and
\begin{equation}
\langle \rho(\qq,\omega)\rangle = [N\rho_0(\omega)+N_i\delta\rho(\qq,\omega)]
\delta_{\qq,0}.
\end{equation}
In other words, one does not simply find that $\rho(\qq,\omega)$ is
broadened by disorder, but rather that in the process of configuration
averaging $\rho(\qq,\omega)$, one loses the finite-$\qq$ signal
altogether.

It is therefore the goal of this paper to develop a systematic
disorder-averaged
formalism for understanding many-impurity effects in FTSTS.  In
Sec.~\ref{simple_model}, we demonstrate that the variance, $\langle
\Delta \rho^2(\qq,\omega) \rangle \equiv
\langle|\rho(\qq,\omega)|^2\rangle-|\langle\rho(\qq,\omega)\rangle|^2$,
retains the useful finite-$\qq$ information when disorder-averaged,
and that the random noise described by $f(\qq)$ is eliminated as a
result of configuration averaging.  Results are illustrated with
numerical calculations described in Sec.~\ref{numerics}.  Finally, we
show in Sec.~\ref{diagrammatic} that one can develop a diagrammatic
perturbation theory for $\langle \Delta\rho^2(\qq,\omega)\rangle $ in
which many-impurity effects can be understood systematically.  One
useful consequence of these calculations is a demonstration that the
mean free path can be extracted from FTSTS spectra.  A summary of
these results will be presented in Sec.~\ref{summary}.

\section{Theory and Calculations}
\label{theory}

\subsection{Simple Model}
\label{simple_model}
For a finite number $N_\textrm{cfg}$ of impurity configurations, the
variance of the FTDOS is $\langle \Delta\rho^2(\qq,\omega)
\rangle_{N_\textrm{cfg}} \equiv
\langle|\rho(\qq,\omega)|^2\rangle_{N_\textrm{cfg}} -
|\langle\rho(\qq,\omega)\rangle_{N_\textrm{cfg}}|^2$.  We keep
$N_\textrm{cfg}$ finite for the moment because, in practical
applications where it can be time consuming to obtain spectra of the
disordered system, one might find that it is only possible to average
over a relatively small number of configurations, and it is therefore
useful to understand how the noise scales with $N_\textrm{cfg}$.
Configuration averaging is performed by averaging over sets of
impurity positions denoted by $\RR^n_i$ with $n =
1,\ldots,N_\textrm{cfg}$ indicating the different configurations, and
$i=1,\ldots,N_i$ labelling impurities.
We typically need to compute averages of the form
\[
\left \langle \sum_{i=1}^{N_i} e^{-i\qq\cdot\RR_i} \right
\rangle_{N_\textrm{cfg}} \equiv
\frac{1}{N_\textrm{cfg}}\sum_{n=1}^{N_\textrm{cfg}} \sum_{i=1}^{N_i}
e^{-i\qq\cdot\RR^n_i} .
\]
Since, for fixed $\qq$, this is just a random walk in the complex plane with
$N_\textrm{cfg}N_i$ steps, it follows that 
\begin{eqnarray}
\left \langle \sum_{i=1}^{N_i} e^{-i\qq\cdot\RR_i} \right
\rangle_{N_\textrm{cfg}} 
 &=& N_i \delta_{\qq,0} +
%\delta_{\qq\neq 0} 
\sqrt{\frac{N_i}{N_\textrm{cfg}}} f(\qq),
\label{cavg}
\end{eqnarray}
where, as always, $f(\qq)$ is a random complex function of $\qq$ for
$\qq\neq 0$ with $|f(\qq)| \sim O(1)$ and $f(0)=0$.  Using
Eqs.~(\ref{ftdos1}) and (\ref{cavg}), one can show that
\begin{eqnarray}
\langle \rho(\qq,\omega)\rangle_{N_\textrm{cfg}} &=& 
[N \rho_0(\omega) + N_i \delta \rho(\qq,\omega)] \delta_{\qq,0}
\nonumber \\
&&\qquad +%\delta_{\qq\neq 0} 
\delta\rho(\qq,\omega) \sqrt{\frac{N_i}{N_\textrm{cfg}}}f(\qq),
\label{ftdos2}
\end{eqnarray}
and that 
\begin{eqnarray}
\langle |\rho(\qq,\omega)|^2\rangle_{N_\textrm{cfg}} &=& [N
\rho_0(\omega) + N_i \delta \rho(\qq,\omega)]^2 \delta_{\qq,0}
\nonumber\\ 
&&+  %\delta_{\qq\neq 0}
 N_i| \delta\rho(\qq,\omega)|^2
\left[ 1 + \frac{\mbox{Re}f(\qq)}{\sqrt{N_\textrm{cfg}}} \right ],
\label{ftdos3}
\end{eqnarray}
so that 
\begin{equation}
\langle \Delta\rho^2(\qq,\omega)\rangle_{N_\textrm{cfg}} = 
N_i|\delta\rho(\qq,\omega)|^2 \left[ 1 +
\frac{\mbox{Re}f(\qq)}{\sqrt{N_\textrm{cfg}}} \right ],
\label{ftdos4}
\end{equation}
to leading order in $N_\textrm{cfg}$.
Equation (\ref{ftdos4}) is the principal result for this subsection,
and it demonstrates the key ideas of this paper which are that (i) the
signal-to-noise ratio of the variance of $\rho(\qq,\omega)$ {\em
improves} with configuration averaging and (ii)
$\lim_{N_\textrm{cfg}\rightarrow\infty} \langle \Delta
\rho^2(\qq,\omega)\rangle_{N_\textrm{cfg}}$ is nonvanishing.  The
latter point suggests that one can construct a disorder-averaged
perturbation series for $\langle \Delta \rho^2(\qq,\omega)\rangle$,
which we will do in Sec.~\ref{diagrammatic}.

\begin{figure}[tb]
\includegraphics[width=\columnwidth]{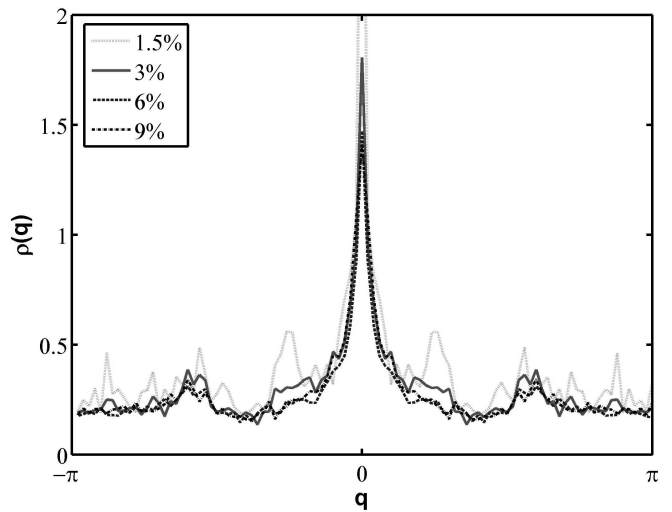}
\caption{(color online) Scaling of FTDOS with number of impurities.  Cuts of
$\rho_\textrm{rms}(\qq,\omega)/\sqrt{N_i}$ along $\qq = (q,0)$ are
shown for different impurity concentrations and for
$N_\textrm{cfg}=29$, $\omega=0$.  
%{\em Inset:} Detail of the FTDOS near $\qq_1$.
}
\label{fig2}
\end{figure}

\subsection{Numerical Method}
\label{numerics}
In order to illustrate the results of the previous section, and to
compare with the diagrammatic perturbation series presented in the
next section, we have calculated the FTDOS exactly for noninteracting
quasiparticles in a disordered 2D metal.  For simplicity, we take
pointlike disorder and a tight-binding band structure of the form
\[
\epsilon_\kk = t_0-2t_1(\cos k_x+\cos k_y)+4t_2 \cos k_x \cos k_y.
\]
This band structure is often used to describe high temperature
superconductors, but is chosen  primarily because it is amenable
to the real-space numerical method described here.  For concreteness,
we measure energies in units of $t_1$ and take $t_0=0.5t_1$ and
$t_2=0.45t_1$.  The Fermi surface for this band structure is shown in
Fig.~\ref{fig1}(b).

In real space, the Hamiltonian is defined on a square
lattice
\begin{equation}
\hat H = \sum_{i,j} \sum_\sigma t_{ij}
c^\dagger_{\rr_i\sigma}c_{\rr_j\sigma} + V_0 \sum_{\ell=1}^{N_i}\sum_{\sigma}
c^\dagger_{\RR_\ell \sigma} c_{\RR_\ell \sigma}
\label{ham}
\end{equation}
where $c_{\rr_i\sigma}$ is the annihilation operator for a
quasiparticle on site $i$ with spin $\sigma$, and $t_{ii} = t_0$,
$t_{ij} = -t_1$ for $i$ and $j$ nearest neighbor sites, and
$t_{ij}=t_2$ for next-nearest-neighbor sites.  As before, $\RR_\ell$
are the impurity site locations, and the impurities have strength
$V_0=10t_1$. Equation (\ref{ham}) can be diagonalized numerically, and the
eigenenergies $E_n$ and eigenvectors $\Psi_n(\rr_i)$ can be inserted
into Eq.~(\ref{exactldos}) to find $\rho(\rr,\omega)$, from which
$\Delta\rho^2(\qq,\omega)$ follows directly.  We remark that for
simulations in which 
 $N_\textrm{cfg}=1$, we plot $|\rho(\qq,\omega)|$, while for
$N_\textrm{cfg}>1$, we actually plot the root-mean-square variance
\[
\rho_\textrm{rms}(\qq,\omega) \equiv
\sqrt{\langle \Delta \rho^2(\qq,\omega) \rangle_{N_\textrm{cfg}}},
\]
which is most the most natural quantity to compare with $|\rho(\qq,\omega)|$.

The numerical approach adopted here restricts the system size to
 $\lesssim 10000$ lattice sites, and other methods\cite{othertmat}
 exist which allow one to study larger systems in the dilute impurity
 limit.  However, in many interesting situations (eg.~ones in which
 interactions are important) numerical calculations will be similarly
 restricted. This system size is also comparable to typical surface
 areas studied in FTSTS experiments.  In general, one can expect the
 FTDOS to be consistent with the predictions of the perturbation
 theory introduced below provided the system is larger than the
 mean-free path.

The scaling with $N_i$ and $N_\textrm{cfg}$ suggested by
Eq.~(\ref{ftdos4}) has been tested numerically using the method
outlined in this section and holds for a wide range of impurity
concentrations, as shown in Fig.~\ref{fig1} and Fig.~\ref{fig2}. In
Fig.~\ref{fig1}, one sees that an average over only four
configurations improves the signal-to-noise ratio significantly (by
about a factor of 2) and that fine details of the FTDOS can be
resolved when one averages over $\approx 10$ configurations.  One also
sees in Fig.~\ref{fig2} that the spectrum generally scales as
$\sqrt{N_i}$ as expected.  Closer inspection reveals that most of the
peaks evident at smaller impurity concentrations are due to finite size
effects and disappear at larger
concentrations where the mean-free-path is less than the system
size. Furthermore, the peak at $\qq\approx (1.9,0)$, corresponding to
$\qq_1$ in Fig.~\ref{fig1}, deviates from the simple scaling at
smaller values of $N_i$, but satisfies the $\sqrt{N_i}$-scaling
relation best when the mean-free-path is less than the system size.

\subsection{Diagrammatic Method}
\label{diagrammatic}
\begin{figure}[tb]
\includegraphics[width=\columnwidth]{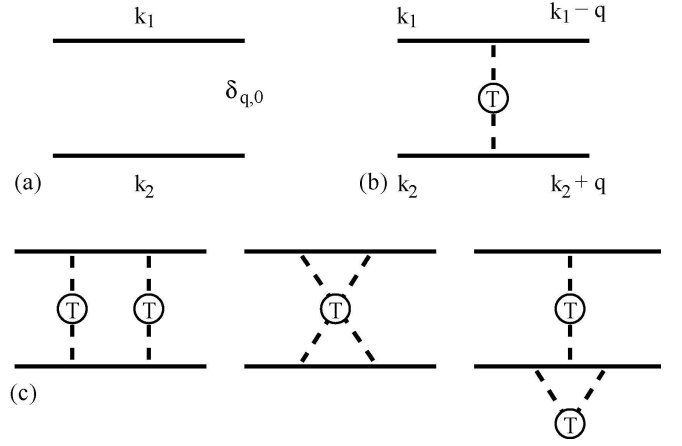}
\caption{Diagrams relevant to the calculation of $\langle \Delta
\rho^2(\qq,\omega) \rangle$.  (a) Disconnected diagrams cancel in the
difference $\langle |\rho(\qq,\omega)|^2\rangle -|\langle
\rho(\qq,\omega) \rangle |^2$.  (b) Leading order and (c) second order
diagrams which are computed in the text.  In all diagrams, solid lines
represent fully self-consistent Green's functions (retarded or
advanced) and each pair of dashed lines connecting to a vertex labelled
``T'' represents $n_i T^{\pm}(\omega) T^{\pm}(\omega)$.}
\label{fig3}
\end{figure}

The goal of this section is to develop a diagrammatic
perturbation series for $\langle \Delta \rho^2(\qq,\omega)\rangle$.
The LDOS is defined in terms of single-particle Green's functions
as
\[
\rho(\rr,\omega) = -\frac{1}{2\pi i}\left [ {\cal G}^+(\rr,\rr,\omega)
-{\cal G}^-(\rr,\rr,\omega) \right ],
\]
where $\cal G^+$ and $\cal G^-$ are the retarded and
advanced Green's functions for a {\em particular} disorder
configuration.  
%To this end, it is useful to
%define two-particle propagators
%\[
%{\cal I}^{s_1s_2}(\rr_1,\rr_2,\omega) = {\cal G}^{s_1}(\rr_1,\rr_1,\omega)
%{\cal G}^{s_2}(\rr_2,\rr_2,\omega),
%\]
%where $s_1,s_2 = \pm$ such that
Then we can write,
\begin{eqnarray*}
|\rho(\qq,\omega)|^2 &=& \frac{-1}{(2\pi)^2} 
\sum_{s_1,s_2=\pm} 
s_1s_2 \sum_{\rr_1,\rr_2} e^{-i\qq\cdot\rr_{12}} \\
%{\cal I}^{s_1s_2}(\rr_1,\rr_2,\omega).
&&\times{\cal G}^{s_1}(\rr_1,\rr_1,\omega) 
{\cal G}^{s_2}(\rr_2,\rr_2,\omega),\\
&=&\frac{-1}{(2\pi)^2} 
\sum_{s_1,s_2=\pm} s_1s_2\sum_{\kk_1,\kk_2}
   \\
&&\times {\cal G}^{s_1}(\kk_1,\kk_1-\qq,\omega) 
{\cal G}^{s_2}(\kk_2,\kk_2+\qq,\omega),
\end{eqnarray*}
where $\rr_{12} = \rr_1-\rr_2$.  The disorder average
of ${\cal G}^{\pm}$ is 
\[
\langle {\cal G}^{\pm}(\kk,\kk+\qq,\omega) \rangle = {
G}^\pm(\kk,\omega)\delta_{\qq,0},
\]
where $G(\kk,\omega)$ is the usual self-consistent disorder-averaged
Green's function.  It follows that
\begin{eqnarray*}
\langle|\rho(\qq,\omega)|^2 \rangle &=& \left [ \frac{1}{2\pi}
    \sum_{s,\kk} s {\bf G}^s(\kk,\omega) \right ]^2\delta_{\qq,0} \\
    &&+ \frac{1}{(2\pi)^2} \sum_{s_1,s_2=\pm} s_1s_2\sum_{\kk_1,\kk_2}\\
    &&\times\langle \langle{\cal G}^{s_1}(\kk_1,\kk_1-\qq,\omega) 
    {\cal G}^{s_2}(\kk_2,\kk_2+\qq,\omega) \rangle\rangle,
\end{eqnarray*}
and $\langle\langle \ldots \rangle\rangle$ refers to a
disorder-averaging process in which only {\em connected} diagrams are
retained.  The first term on the right hand side of this equation
corresponds to the disconnected diagram shown in Fig.~\ref{fig3}(a),
and is also equal to $|\langle \rho(\qq,\omega)\rangle |^2$.  Thus,
the variance of $\rho(\qq,\omega)$ is
\begin{equation}
\langle\Delta \rho^2(\qq,\omega)\rangle = \frac{-1}{(2\pi)^2}
\sum_{s_1,s_2=\pm} s_1s_2{ I}^{s_1s_2}(\qq,\omega)
\label{dr}
\end{equation}
where the disorder-averaged two-particle response kernel 
${I}^{s_1s_2}(\qq,\omega)$
is
\begin{equation}
I = \frac{1}{N^2} \sum_{\kk_1,\kk_2}
\langle\langle {\cal G}^{s_1}(\kk_1,\kk_1-\qq,\omega) {\cal
G}^{s_2}(\kk_2,\kk_2+\qq,\omega) \rangle\rangle.
\label{GG}
\end{equation}

We first consider the leading order contribution to the FTDOS, which
is shown in Fig.~\ref{fig3}(b) and which can be written as
\begin{eqnarray}
\langle \Delta \rho^2_1(\qq,\omega)\rangle &=& N_i \Bigg [
\frac{\mbox{Im}}{\pi N} \sum_\kk G^{+}(\kk,\omega)
T^+_{\kk,\kk+\qq}(\omega)\nonumber \\ &&\times G^{+}(\kk+\qq,\omega)
\Bigg ]^2,
\label{leadingorder}
\end{eqnarray}
where $T^+_{\kk,\kk+\qq}(\omega)$ is the retarded scattering T-matrix
describing scattering from a single impurity embedded in a medium with
impurity concentration $n_i = N_i/N$.  The astute reader will
recognize that this appears to simply be the square of single-impurity
result (multiplied by $N_i$) which is widely used to describe the FTDOS in
strongly-correlated superconductors.\cite{dhlee} The key difference
here is that the Green's functions and T-matrix are now evaluated at
the level of the self-consistent T-matrix approximation\cite{sctma} (SCTMA).
%which is valid in the limit of low impurity concentration.  
This approximation amounts to solving the one-impurity problem in an
effective medium characterized by a self-energy $\Sigma(\kk,\omega)$
which approximately describes the remaining $N_i-1$ impurities.  In
this approximation, the self-consistency condition is
$\Sigma(\kk,\omega) = n_iT_{\kk,\kk}(\omega)$, and it can be shown
that this is equivalent to the coherent potential approximation in the
dilute limit. This result---that the addition of a self-energy to the
Green's function accounts for many-impurity effects to leading
order---has been taken as an ansatz by several
authors,\cite{ansatz1,ansatz2} but has not previously been formally
justified.  

\begin{figure}[tb]
\includegraphics[width=\columnwidth]{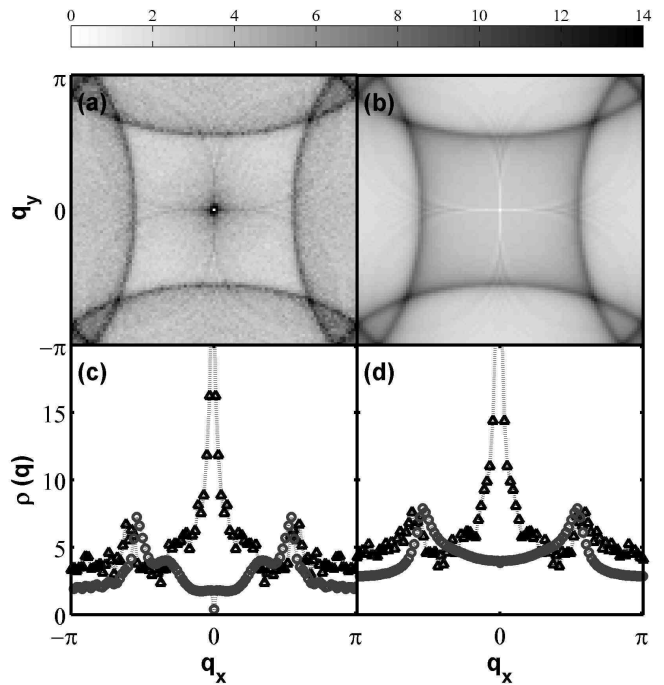}
\caption{(color online) Comparison of diagrammatic and numerical
$\rho_\textrm{rms}(\qq,\omega)/\sqrt{N_i}$ at $\omega=0$. Results are
shown for 3\% impurities for (a) the numerical method with
$N_\textrm{cfg}=29$, and (b) the leading order diagrammatic result
given by Eq.~(\protect\ref{leadingorder}).  Cuts along the line $\qq =
(q,0)$ are shown in (c) and (d) for 3\% and 6\% disorder respectively
for exact (red triangles) and diagrammatic (blue circles) calculations.}
\label{fig4}
\end{figure}

The evaluation of Eq.~(\ref{leadingorder}) is simplest in the case of
a random distribution of pointlike impurities with scattering
potential potential $V_0$.  In this case, the T-matrix loses its
dependence on $\kk$ and $\qq$ and we have the following system of
equations which must be self-consistently solved:
\begin{eqnarray*}
\Sigma^{\pm}(\omega) &=& n_i  T^{\pm}(\omega) \\ 
 G^\pm(\kk,\omega) &=&
[\omega^\pm - \epsilon_\kk - \Sigma^\pm(\omega)]^{-1} \\
g^{\pm}(\omega) &=& \frac{1}{N}\sum_\kk G^\pm(\kk,\omega) \\
T^\pm(\omega) &=& V_0/[1 - V_0 g^{\pm}(\omega) ]
\end{eqnarray*}
with $\omega^\pm = \omega \pm i\eta$ and $\eta$ a positive, real,
infinitessimal. The results of this calculation are compared with
numerical results of the previous section in Fig.~\ref{fig4}.  In
general, the agreement between the leading-order diagrammatic and
numerical approaches is good, except for the broad peak centered at
$\qq = 0$ which is present in the numerical calculations and absent in
the leading-order approximation.  This peak is distinct from the
$\qq=0$ peak in Eq.~(\ref{ftdos1}) (which occurs strictly at $\qq=0$
and is $\sim O(N)$) and arises from higher order diagrams.  The
finite-q peak heights in the diagrammatic calculations scale as
$\sqrt{N_i}$, as shown in Fig.~\ref{fig4}(c) and (d), in agreement
with the scaling of the numerical results discussed previously.  The
scattering rate in (d) is approximately twice that in (c), and it is
interesting that, apart from the overall scaling of the spectrum, the
most significant effect of scattering is to enhance the background.
Thus, one's ability to resolve the peaks is not limited by a reduction
of their height, but by a reduction in contrast with the background.

\begin{figure}[tb]
\includegraphics[width=\columnwidth]{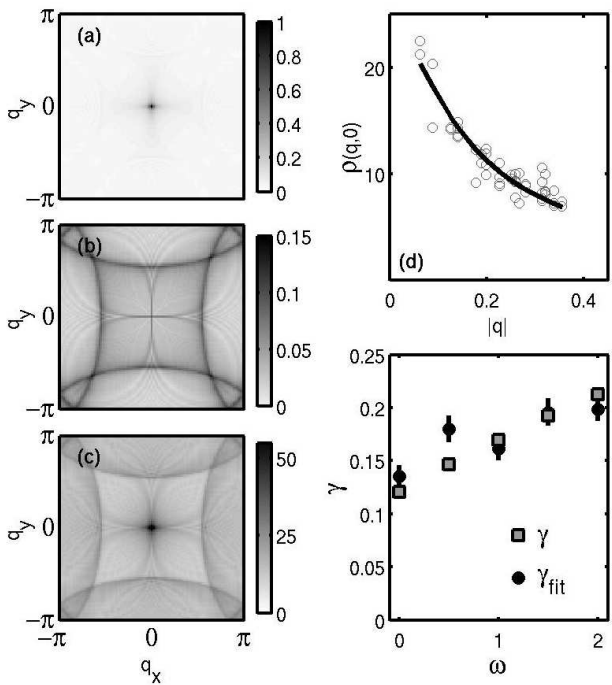}
\caption{(color online) Second order corrections to
$\rho_\textrm{rms}(\qq,\omega)$.  Panels (a) and (b) show the vertex
functions $W^{+-}(\qq,\omega)$ and $W^{++}(\qq,\omega)$ respectively
from which second order corrections are calculated, (c) shows
$\rho_\textrm{rms}(\qq,\omega)$ evaluated using
Eq.~(\protect\ref{I2}); (d) shows a least-squares fit of
Eq.~(\protect\ref{ell}) to the results of exact numerical calculations
at $\omega=0$;
(e) shows the scattering rate $\gamma$ extracted from the SCTMA
self-energy
%$\gamma=-\mbox{Im } \Sigma^+(\omega)$ 
(square) and from fits of Eq.(\protect\ref{ell}) (solid circle) to the
$\qq=0$ peak of numerical calculations.  Error bars are the standard
deviation between the fitted curve and data.  Panels (a)-(c) are at
$\omega = 0$ and for $n_i=0.03$, while (d) and (e) are for $n_i=0.06$ and
$N_\textrm{cfg}=29$.}
\label{fig5}
\end{figure}

In order to get an understanding of the effects of higher order
diagrams, we consider the three possible second order diagrams shown
in Fig.~\ref{fig3}(c), which combine to give
\begin{eqnarray}
I_{2}^{s_1s_2}(\qq) &=& \frac{N_iT^{s_1}T^{s_2}}{N^2}
\sum_{\kk_1,\kk_2} G^{s_1}(\kk_1 ) G^{s_1}(\kk_1-\qq ) \nonumber \\
&&\times G^{s_2}(\kk_2 ) G^{s_2}(\kk_2+\qq ) [W^{s_1s_2}(\kk_1+\kk_2
)\nonumber \\ &&+ W^{s_1s_2}(\kk_1-\kk_2-\qq) + 2W^{s_1s_1}(\qq)]
\\[3mm] W^{s_1s_2}({\bf p} ) &=& n_i T^{s_1} T^{s_2} \frac 1 N
\sum_\kk G^{s_1}(\kk)G^{s_2}(\kk+{\bf p}), \label{Wpm}
\end{eqnarray}
where we note that $W^{++}(\pp) = {W^{--}(\pp)}^\ast$ and $W^{+-}(\pp)
= W^{-+}(\pp)$.  Both $W^{++}(\pp)$ and $W^{+-}(\pp)$ are shown in
Fig.~\ref{fig5} at $\omega=0$.  The term $W^{+-}(\pp) \approx 1-
Dp^2/2\gamma$ for small $p$, where $D$ is the diffusion coefficient of
the metal,\cite{wl} and is substantially larger than $W^{++}(\pp)$.

We are interested in the largest contribution from the second order diagrams,
which comes from the retarded-advanced ($W^{+-}$) channel.  Because
$W^{+-}(\pp)$ is peaked at $\pp=0$, we can evaluate Eq.~(\ref{dr}) 
approximately as:
\begin{eqnarray}
\langle \Delta \rho^2_2(\qq,\omega) \rangle &=& 
\langle \Delta \rho^2_1(\qq,\omega) \rangle +
\frac{N_iT^+T^-}{\pi^2N^2}  \sum_{\pp} 
W^{+-}(\pp)\nonumber  \\
&&\times \sum_{\kk}
|G^+(\kk) G^+(\kk+\qq)|^2.
 \label{I2} 
 \end{eqnarray}
 The results of this calculation are shown in Fig.~\ref{fig5}(c).  One
 sees a weak background structure which is essentially the same as
 that from the leading order contribution in Eq.~(\ref{leadingorder}),
 but the dominant feature in the second order correction is the
 central peak.  Physically, this suggests that two-impurity processes
 generally interfere destructively and tend to shift weight to small
 values of $\qq$.  At this level of approximation, however, we do not
 capture the physics of trapped states, which have been predicted to
 occur in d-wave superconductors between pairs of impurities
 satisfying an appropriate resonance criterion.\cite{zhu2imp} Such
 resonances would tend to enhance finite-$\qq$ peaks in the FTDOS.
 
We note that, while Fig.~\ref{fig5}(c) has the same qualitative
structure as the exact numerical results shown in Fig.~(\ref{fig4}),
the quantitative agreement is still poor.  This is because,  
in the absence of inelastic dephasing or coupling to the
third dimension, weak localization physics affects the FTDOS.
One can attempt to go beyond second order in the perturbation series,
and the natural step to take is to sum either the ladder or maximally
crossed diagrams.  Both of these calculations give the same result:
\[
\langle \Delta \rho^2(\qq,\omega) \rangle \sim \sum_\pp \frac{1}{Dp^2}
\sum_\kk |G^+(\kk) G^+(\kk+\qq)|^2.
\]
In 2D, the sum over $\pp$ diverges logarithmically, but the divergence
will be cut
off either by inelastic dephasing or by tunneling out of the surface
states.\cite{wl} What is perhaps most interesting about this result,
however, is the $\qq$-space structure of the central peak, which is the
same as in Eq.~(\ref{I2}).
 
 We can estimate the form of the central peak in $\langle \Delta
 \rho^2(\qq,\omega)\rangle$ in the limit $q^2/2m \ll v_Fq$, where
for an isotropic Fermi surface:
 \begin{equation}
 \frac{1}{N} \sum_{\kk} |G^+(\kk) G^+(\kk+\qq)|^2
 = \frac{\pi N_0}{\gamma^2} \frac{1}{\sqrt{(v_Fq)^2+(2\gamma)^2}}, 
 \end{equation}
such that, for small $\qq$,
 \begin{equation}
 \langle \Delta\rho^2(\qq,\omega)\rangle = \frac{A}{\sqrt{1+(q\ell/2)^2}}
\label{ell}
 \end{equation}
  where $\ell = v_F/\gamma$ is the mean-free path and $A$ dictates the
peak weight.  This result suggests that one can extract $\ell$ from
FTSTS experiments.  As a check, we have made least-squares fits of
Eq.~(\ref{ell}) to the central peak of an exact numerical calculation
of the type outlined in Sec.~\ref{numerics} with $n_i = 0.06$ and
$N_\textrm{cfg}=29$. 
%
% CHANGED IN RESPONSE TO REFEREE REPORT:
We have treated both $A$ and $\ell$ as fitting
parameters, and an example of the numerical data
and fitted curve shown in Fig.~\ref{fig5}(d) 
illustrates that Eq.~(\ref{ell}) appears to describe the
exact result rather well.  
In Fig.~\ref{fig5}(e), the scattering
rate $\gamma = v_F/\ell$ extracted from the fitting procedure is shown
as a function of $\omega$. 
% END CHANGES
For comparison the scattering rate $\gamma =
-\mbox{Im }\Sigma^+(\omega)$ calculated using the SCTMA is also shown.
%
% CHANGED IN RESPONSE TO REFEREE REPORT:
%An example of the data and best-fit at $\omega = 0$ is shown in the
%inset, and shows that the Eq.~(\ref{ell}) appears to describe the
%exact result rather well.  
Because the band structure is anisotropic, we have chosen to fit
$\rho(\qq,\omega)$ as a function of the radial coordinate $q=|\qq|$
(note that the point at $\qq=0$ must be excluded), as shown in
Fig.~\ref{fig5}(d).  The fit is therefore to an angular average of the
data, and the error bars in Fig.~\ref{fig5}(e) are the
root-mean-square deviation between the fit and the data.  We have
experimented with fitting to cuts through the data along various
directions, for example $\qq=(q,0)$ and $\qq=(q,q)$, but have
generally found this to be less reliable than fitting to the angular
average.  In general, 
%we have found that the fitting procedure is made
%difficult because uniqueness of the fitting parameters requires a good
%knowledge of $\rho(\qq,\omega)$ for $q < \ell^{-1}$, and noise in the
%$\qq=0$ point has a disproportionate influence on the value of $\ell$.
%
% CHANGED IN RESPONSE TO REFEREE REPORT:
since we are attempting to fit a peak of half-width $\approx
2\ell^{-1}$, a reliable fit requires a q-space resolution $\Delta q
\lesssim 2\ell^{-1}$,
% The fitting
%procedure works best when the q-space resolution $\Delta q$ is much
%less than $\ell^{-1}$, 
or (equivalently) 
%when 
that the region over which one makes the LDOS map is
%much 
larger than roughly twice the mean-free-path.  With
these caveats in mind, the good agreement between the self-consistent
T-matrix and exact numerical results is heartening, and shows that one
can legitimately hope to extract the mean-free path from experiments.
 
% ADDED IN RESPONSE TO REFEREE REPORT:
Throughout this work, we have presumed that one is averaging distinct
samples.  However, for experimental purposes, it may be more
straightforward to measure a single large surface area (of side $L$),
and partition the data into $n^2$ subregions of side $L/n$.  The
benefits of this procedure are that one gains a factor of $n$ in the
signal-to-noise ratio, but the cost is a loss of q-space resolution by
the same factor $n$.  Provided $L/n \gtrsim \ell$, however, one is not
losing important information as a result of the partitioning process.

We finish this section with a brief discussion of inelastic
scattering.  The essential point is that, since the correlation
function measured by $\Delta \rho^2(\qq,\omega)$ is between
single-particle Green's functions, and is not a two-particle
correlation function, there are no ``vertex-correction'' diagrams in
which inelastic propagators connect the upper and lower Green's
functions in Fig.~\ref{fig3}.  To put it another way, unless one has
localized bosonic modes attached to the impurity sites, inelastic
scattering will be spatially uniform and will not produce Friedel
oscillations.  Such Friedel oscillations would be described by
inelastic vertex-correction diagrams, and thus these diagrams must be
absent.  The effect of inelastic scattering is, therefore, solely to
modify the single-particle self-energy.
%If Matthiessen's rule applies, the total
%self-energy is $\Sigma(\omega) = \Sigma_\textrm{el}(\omega) +
%\Sigma_\textrm{inel}(\omega)$ where $\Sigma_\textrm{el}$ and
%$\Sigma_\textrm{inel}$ are the elastic and inelastic self-energies
%respectively.  
Since it is the total scattering rate which determines
peak heights and widths in $\Delta \rho^2(\qq,\omega)$, separation of
elastic and inelastic scattering rates is difficult.  In particular,
the scattering rate one extracts from Eq.~(\ref{ell}) contains both
the elastic and inelastic scattering rates.

\section{Summary}
\label{summary}
We have studied the disorder-averaged variance $\langle \Delta
\rho^2(\qq,\omega) \rangle$ of the density of states of a disordered
2D metallic surface.  Through a combination of numerical and analytic
calculations, we have shown that disorder-averaging is an effective
means of increasing the signal-to-noise ratio in the Fourier
transformed local density of states.  This resolves a problem with the
conventional experimental technique of measuring $\rho(\qq,\omega)$
for which the signal-to-noise ratio remains fixed as one averages over
impurity configurations.  Furthermore, we have shown that one can
construct a diagrammatic perturbation theory for $\langle \Delta
\rho^2(\qq,\omega) \rangle$ and that it is possible to extract an
important physical property, the mean-free-path, from the central peak
in the FTSTS spectrum.


\begin{thebibliography}{stuffstuffstuff}
\bibitem{othertmat} One can solve directly for the many-impurity T-matrix
as in Ref.~\protect\onlinecite{ZAH}.
\end{thebibliography}
\end{document}